\documentclass{PoS}
\usepackage{amssymb}
\usepackage{amsmath}
\usepackage{mathtools}
\usepackage{subfig}
\usepackage{graphicx}
\usepackage{placeins}
\usepackage{bm}
\usepackage{multirow}

\title{NLO and NNLO Low Energy Constants for $\bm{SU(2)}$ Chiral Perturbation Theory}

\ShortTitle{NLO and NNLO LECs for $SU(2)~\chi \mathrm{PT}$}

\author{R.D.~Mawhinney and \speaker{D.J.~Murphy} for the RBC-UKQCD Collaboration\footnote{The numerical results of QCD simulations used in these fits were produced on the QCDOC and BG/Q computers at Brookhaven National Lab and the University of Edinburgh and on the BG/Q computers at the ALCF at Argonne National Lab and Lawrence Livermore National Laboratory. These computers are supported by the U.S. Department of Energy, the RIKEN-BNL Research Center, and the U.K. DiRAC program. We thank J.~Bijnens for Fortran programs implementing the NNLO $\mathrm{PQ} \chi \mathrm{PT}$ formulae.}\\
        Department of Physics, Columbia University, New York, NY 10027, USA\\
        E-mail: \email{rdm@physics.columbia.edu}, \email{dmurphy@phys.columbia.edu}}

\abstract{We have performed global fits of $f_{\pi}$ and $m_{\pi}$, from a variety of RBC-UKQCD domain wall fermion ensembles, to $SU(2)$ partially quenched chiral perturbation theory at NNLO. We report values for 9 NLO and 8 linearly independent combinations of NNLO partially quenched low energy constants, which we compare to other lattice and phenomenological determinations. We discuss the convergence of the expansion and use our large set of low energy constants to make predictions for the pion mass splitting due to QCD isospin breaking effects and the s-wave $\pi \pi$ scattering lengths. We conclude that, for the range of pseudoscalar masses explored in this work, $115~\mathrm{MeV} \lesssim  m_{\rm PS} \lesssim 430~\mathrm{MeV}$, the NNLO $SU(2)$ expansion is quite robust and can fit lattice data with percent-scale accuracy.}

\FullConference{The 33rd International Symposium on Lattice Field Theory\\
		14 -18 July 2015\\
		Kobe International Conference Center, Kobe, Japan}

\begin{document}

\section{Introduction}

The light pseudoscalar mesons of QCD can be understood to arise as pseudo-Goldstone bosons generated by the spontaneous breaking of chiral symmetry. The fact that they are light compared to other hadronic scales --- the proton and neutron, for example --- motivates an effective field theory description, known as chiral perturbation theory ($\chi \mathrm{PT}$). In $SU(2)$ $\chi \mathrm{PT}$ the up and down quark masses are treated as small perturbations away from the $m_{u} = m_{d} = 0$ limit, where the QCD Lagrangian has an exact $SU(2)_{L} \otimes SU(2)_{R}$ chiral symmetry, allowing for explicit calculations of pion physics in terms of $m_{u}$ and $m_{d}$.

The $SU(2)$ $\chi \mathrm{PT}$ Lagrangian is constructed using a general prescription for effective field theories first proposed by Weinberg: one picks a power counting scheme and writes down the most general Lagrangian for the pion consistent with the $SU(2)_{L} \otimes SU(2)_{R}$ symmetry of massless QCD order-by-order. The relative contributions from each such operator appearing in the Lagrangian are parametrized by \textit{a priori} unknown low energy constants (LECs) which must be determined by matching to experiment or to lattice calculations. $SU(2)$ $\chi \mathrm{PT}$ was first explicitly constructed and explored at next-to leading order (NLO) by Gasser and Leutwyler \cite{Gasser:1983yg}. Full calculations of the partially quenched masses and decay constants at NNLO were later performed by Bijnens and L\"{a}hde \cite{bijnens_nnlo_pq}, who also provided Fortran routines we use to compute the NNLO corrections in our fits.

In this talk we consider fits of RBC/UKQCD lattice data for the light pseudoscalar mesons to the more general $SU(2)$ partially quenched chiral perturbation theory ($\mathrm{PQ} \chi \mathrm{PT}$) at next-to-next-to leading order. In addition to determining low energy constants, the fits allow us to systematically study the behavior and range of applicability of the $SU(2)$ $\mathrm{PQ} \chi \mathrm{PT}$ expansion up to NNLO. Additional detail regarding this work can be found in Ref.~\cite{su2_chpt_paper}. Analogous fits to NNLO $SU(3)$ $\mathrm{PQ} \chi \mathrm{PT}$ are discussed in an accompanying talk \cite{bob_su3_talk}. 

\section{Lattice Setup}

In this analysis we make use of a number of RBC/UKQCD domain wall fermion ensembles with a wide range of unitary pion masses, physical volumes, and inverse lattice spacings, summarized in Table~\ref{tab:ensembles_input}. In all cases we work in the isospin symmetric limit of QCD, with two, degenerate dynamical light quark flavors of bare mass $m_{l}$, and a single dynamical heavy flavor of bare mass $m_{h}$ ($N_{f} = 2 + 1$). We use the Iwasaki gauge action (I), and on some ensembles supplement this with the dislocation suppressing determinant ratio (I+DSDR). We simulate QCD with $N_{f} = 2+1$ quark flavors using the domain wall fermion formalism, with either the Shamir (DWF) or M\"{o}bius (MDWF) kernel. Additional details of the ensemble generation and fits to extract the low-energy QCD spectrum can be found in Refs.~\cite{su2_chpt_paper,Allton:2008pn,Aoki:2010dy,Arthur:2012opa,Blum:2014tka}.

The fits we have performed include data for the pion mass and decay constant, the kaon mass and decay constant, and the $\Omega$ baryon mass on each ensemble. On the older 24I, 32I, and 32ID ensembles these measurements were performed for a number of different partially quenched valence quark mass combinations. In addition, reweighting in the dynamical heavy quark mass was used to determine the $m_{h}$ dependence and allow for a small, linear interpolation from the simulated $m_{h}$ to the physical value. On the newer ensembles --- 32I-fine, 48I, 64I, 32ID-M1, and 32ID-M2 --- we perform a single set of unitary measurements of the same observables, and do not reweight in $m_{h}$.

\begin{table}[h]
\centering
\resizebox{\columnwidth}{!}{
\begin{tabular}{c||c|cccc|cc|c}
\hline
\hline
\rule{0cm}{0.4cm} Ensemble & Action & $\beta$ & $L^{3} \times T \times L_{s}$ & $a m_{l}$ & $a m_{h}$ & $m_{\pi} L$ & $m_{\pi}$ (MeV) & $a^{-1}$ (GeV) \\
\hline
\rule{0cm}{0.4cm} \multirow{2}{*}{24I} & DWF+I & 2.13 & $24^{3} \times 64 \times 16$ & 0.005 & 0.04 & 4.57(1) & 339.6(1.2) & \multirow{2}{*}{1.784(5)} \\
\rule{0cm}{0.4cm} & DWF+I & 2.13 & $24^{3} \times 64 \times 16$ & 0.01 & 0.04 & 5.81(1) & 432.2(1.4) & \\
\hline
\rule{0cm}{0.4cm} \multirow{3}{*}{32I} & DWF+I & 2.25 & $32^{3} \times 64 \times 16$ & 0.004 & 0.03 & 4.06(1) & 302.0(1.1) & \multirow{3}{*}{2.382(8)} \\
\rule{0cm}{0.4cm} & DWF+I & 2.25 & $32^{3} \times 64 \times 16$ & 0.006 & 0.03 & 4.838(8) & 359.7(1.2) & \\
\rule{0cm}{0.4cm} & DWF+I & 2.25 & $32^{3} \times 64 \times 16$ & 0.008 & 0.03 & 5.53(1) & 410.8(1.5) & \\
\hline
\rule{0cm}{0.4cm} \multirow{2}{*}{32ID} & DWF+I+DSDR & 1.75 & $32^{3} \times 64 \times 32$ & 0.001 & 0.046 & 3.999(7) & 172.7(9) & \multirow{2}{*}{1.378(7)} \\
\rule{0cm}{0.4cm} & DWF+I+DSDR & 1.75 & $32^{3} \times 64 \times 32$ & 0.0042 & 0.046 & 5.792(8) & 250.1(1.2) & \\
\hline
\rule{0cm}{0.4cm} 32I-fine & DWF+I & 2.37 & $32^{3} \times 64 \times 12$ & 0.0047 & 0.0186 & 3.77(4) & 370.1(4.4) & 3.144(17) \\
\hline
\rule{0cm}{0.4cm} 48I & MDWF+I & 2.13 & $48^{3} \times 96 \times 24$ & 0.00078 & 0.0362 & 3.863(6) & 139.1(4) & 1.729(4) \\
\hline
\rule{0cm}{0.4cm} 64I & MDWF+I & 2.25 & $64^{3} \times 128 \times 12$ & 0.000678 & 0.02661 & 3.778(8) & 139.0(5) & 2.357(7) \\
\hline
\rule{0cm}{0.4cm} 32ID-M1 & MDWF+I+DSDR & 1.633 & $32^{3} \times 64 \times 24$ & 0.00022 & 0.0596 & 3.78(2) & 117.3(4.4) & 0.981(39) \\
\hline
\rule{0cm}{0.4cm} 32ID-M2 & MDWF+I+DSDR & 1.943 & $32^{3} \times 64 \times 12$ & 0.00478 & 0.03297 & 6.24(2) & 401.0(2.3) & 2.055(11) \\
\hline
\hline
\end{tabular}
}
\caption{Summary of ensembles included in this analysis and input parameters. Here $\beta$ is the gauge coupling, $L^{3} \times T \times L_{s}$ is the lattice volume decomposed into the length of the spatial ($L$), temporal ($T$), and fifth ($L_{s}$) dimensions, and $a m_{l}$ and $a m_{h}$ are the bare, input light and heavy quark masses. The value of $m_{\pi}$ quoted is the unitary pion mass in physical units.}
\label{tab:ensembles_input}
\end{table}

\section{The Global Fit Procedure}

In Ref.~\cite{Aoki:2010dy, Arthur:2012opa, Blum:2014tka} we have developed a ``global fit" procedure for performing a combined chiral fit and continuum extrapolation of lattice data, the details of which we will summarize here. The chiral ans\"{a}tze we use reflect a simultaneous expansion in the quark masses\footnote{We use the notation $m_{x}$ and $m_{y}$ for the valence quarks, and $m_{l}$ and $m_{h}$ for the light and heavy sea quarks.}, lattice spacing, and box size about the chiral, continuum, and infinite-volume limit. The chiral ansatz for each observable has the schematic form
\begin{itemize}
\item $m_{\pi}^{2}$ and $f_{\pi}$: ( NLO or NNLO continuum $SU(2)$ $\mathrm{PQ} \chi \mathrm{PT}$ ) +  ( $\Delta_{\rm FV}^{\rm NLO}$ ) +  ( $c_{a} a^{2}$; for $f_{\pi}$ )
\item $m_{K}^{2}$ and $f_{K}$: ( NLO continuum heavy-meson $SU(2)$ $\mathrm{PQ} \chi \mathrm{PT}$ ) +  ( $\Delta_{\rm FV}^{\rm NLO}$ ) +  ( $c_{a} a^{2}$; for $f_{K}$ )
\item $m_{\Omega}$: ( linear ansatz in $\tilde{m}_{x}$, $\tilde{m}_{l}$, and $\tilde{m}_{h}$ )
\end{itemize}
where $\Delta_{\rm FV}^{\rm NLO}$ denotes NLO finite volume corrections. We work in terms of the total quark masses $\tilde{m}_{q} = m_{q} + m_{\rm res}$, including the residual mass ($m_{\rm res}$) \cite{Allton:2008pn}. We distinguish between ``NLO'' fits, where the continuum $\mathrm{PQ} \chi \mathrm{PT}$ ansatz for the pion is truncated to NLO, and ``NNLO'' fits, where it is truncated to NNLO, and emphasize that it is only the chiral ansatz for the pion which varies. After performing the chiral fit we match to a continuum scaling trajectory by numerically inverting the fit to determine $m_{l}^{\rm phys}$ and $m_{s}^{\rm phys}$ such that the ratios $m_{\pi} / m_{\Omega}$ and $m_{K} / m_{\Omega}$ take their experimentally known values; this defines $m_{\pi}$, $m_{K}$, and $m_{\Omega}$ to have no $\mathcal{O}(a^{2})$ corrections. We then extract the lattice scales $a = m_{\Omega} / m_{\Omega^{-}}^{\rm expt.}$ from the ratio of the simulated $\Omega$ baryon mass on each ensemble, corrected with the chiral fit to $m_{l}^{\rm phys}$ and $m_{s}^{\rm phys}$, to the experimentally determined $\Omega^{-}$ baryon mass. We direct the reader to Ref. \cite{su2_chpt_paper} for additional technical detail regarding the global fit procedure. 

\section{Results}

\begin{figure}[!ht]
\centering
\subfloat{\includegraphics[width=\textwidth]{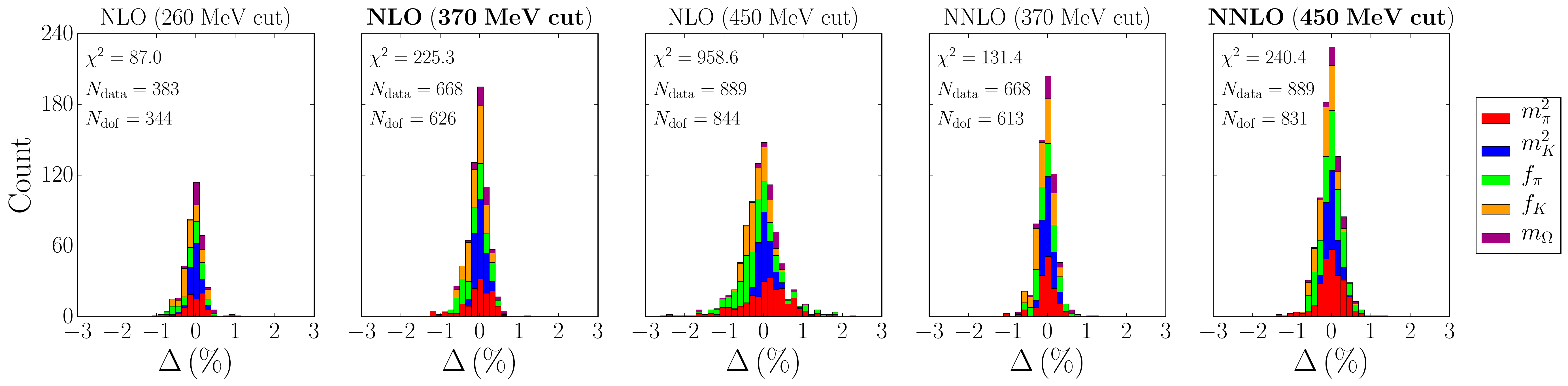}} \\
\subfloat[Summary of LO and NLO LECs]{
	\resizebox{0.5\columnwidth}{!}{
		\begin{tabular}{c|c|c|c}
			\hline
			\hline
			\rule{0cm}{0.4cm}Order & LEC & NLO fit ($370 \, \mathrm{MeV}$ cut) & NNLO fit ($450 \, \mathrm{MeV}$ cut) \\
			\hline
			\rule{0cm}{0.4cm}\multirow{2}{*}{LO} & $B^{\overline{\rm MS}}(\mu = 2 \, \mathrm{GeV})$ & 2.804(34) GeV & 2.787(39) GeV \\
			& $f$ & 121.3(1.5) MeV & 121.5(1.6) MeV \\
			\hline
			\rule{0cm}{0.4cm}\multirow{10}{*}{NLO} & $10^{3} \hat{L}_{0}^{(2)}$ & --- & 1.0(1.1) \\
			& $10^{3} \hat{L}_{1}^{(2)}$ & --- & -0.62(52) \\
			& $10^{3} \hat{L}_{2}^{(2)}$ & --- & 0.06(74) \\
			& $10^{3} \hat{L}_{3}^{(2)}$ & --- & -1.56(87) \\
			& $10^{3} \hat{L}_{4}^{(2)}$ & -0.211(79) & -0.56(22) \\
			& $10^{3} \hat{L}_{5}^{(2)}$ & 0.438(72) & 0.60(28) \\
			& $10^{3} \hat{L}_{6}^{(2)}$ & -0.175(48) & -0.38(10) \\
			& $10^{3} \hat{L}_{7}^{(2)}$ & --- & -0.75(27) \\
			& $10^{3} \hat{L}_{8}^{(2)}$ & 0.594(36) & 0.69(13) \\
			\hline
			\hline
		\end{tabular}
	}
}
\subfloat[Summary of NNLO LECs]{
	\resizebox{0.5\columnwidth}{!}{
		\begin{tabular}{c|c|c}
			\hline
			\hline
			\rule{0cm}{0.4cm}Order & LEC & NNLO fit ($450 \, \mathrm{MeV}$ cut) \\
			\hline
			\rule{0cm}{0.4cm}\multirow{10}{*}{NNLO} & $10^{6} \left( \hat{K}_{17}^{(2)}-\hat{K}_{39}^{(2)} \right)$ & -7.6(1.1) \\
			& $10^{6} \left( \hat{K}_{18}^{(2)}+6\hat{K}_{27}^{(2)}-\hat{K}_{40}^{(2)} \right)$ & 19.2(4.7) \\
			& $10^{6} \hat{K}_{19}^{(2)}$ & -0.9(4.2) \\
			& $10^{6} \hat{K}_{20}^{(2)}$ & -3.2(2.8) \\
			& $10^{6} \left( \hat{K}_{21}^{(2)}+2 \hat{K}_{22}^{(2)} \right)$ & 4.9(4.1) \\
			& $10^{6} \hat{K}_{23}^{(2)}$ & -2.8(1.4) \\
			& $10^{6} \hat{K}_{25}^{(2)}$ & 1.3(1.7) \\
			& $10^{6} \left( \hat{K}_{26}^{(2)}+6\hat{K}_{27}^{(2)} \right)$ & 11.2(3.6) \\
			\hline
			\hline
		\end{tabular}
	}
}
\caption{Top row: stacked histograms of the percent deviation between each data point and the corresponding fit prediction ($\Delta \equiv 200 \times (Y-Y^{\rm fit}) / (Y + Y^{\rm fit})$). Panels (b) and (c): LECs of $SU(2)$ $\mathrm{PQ} \chi \mathrm{PT}$ defined at the chiral scale $\Lambda_{\chi} = 1~\mathrm{GeV}$ from our representative fits. The errors are purely statistical.}
\label{fig:pdev_histograms}
\end{figure}

In Figure \ref{fig:pdev_histograms} we summarize the fits we have performed and their quality with histograms of the percent deviation between each data point and the fit and tables of the values we obtain for the relevant LECs. In all cases we are minimizing an uncorrelated $\chi^{2}$ (see Appendix D of Ref.~\cite{su2_chpt_paper} for more detail regarding correlations in our data). In addition to comparing ``NLO'' and ``NNLO'' fits, we also vary an explicit cut on the heaviest pion mass included in the fit: any ensemble with a unitary pion mass heavier than the cut is excluded completely, as are partially quenched ``pion'' measurements with $m_{x y} > m_{\pi}^{\rm cut}$ on included ensembles. We generally observe excellent consistency between the data and the chiral fits, suggesting that $SU(2)$ $\mathrm{PQ} \chi \mathrm{PT}$ is able to describe our lattice data to $\mathcal{O}(1 \%)$ or better provided the mass cut is chosen appropriately for a given order of the chiral expansion. While we observe that the NLO $SU(2)$ ansatz has clearly started to systematically disagree with the data for the most agressive NLO fit, we also note that the worst outliers are $\mathcal{O}(2-3 \%)$, suggesting that the NLO expansion can still be used with percent-scale accuracy even up to a heavy mass scale $m_{\pi} \sim 450~\mathrm{MeV}$. In the remainder of the talk we focus on the NLO fit with a 370 MeV cut and the NNLO fit with a 450 MeV cut as representative NLO and NNLO fits of good quality.

\begin{figure}[!ht]
\vspace{-0.6cm}
\centering
\subfloat[NLO, 370 MeV cut]{\includegraphics[width=0.32\textwidth]{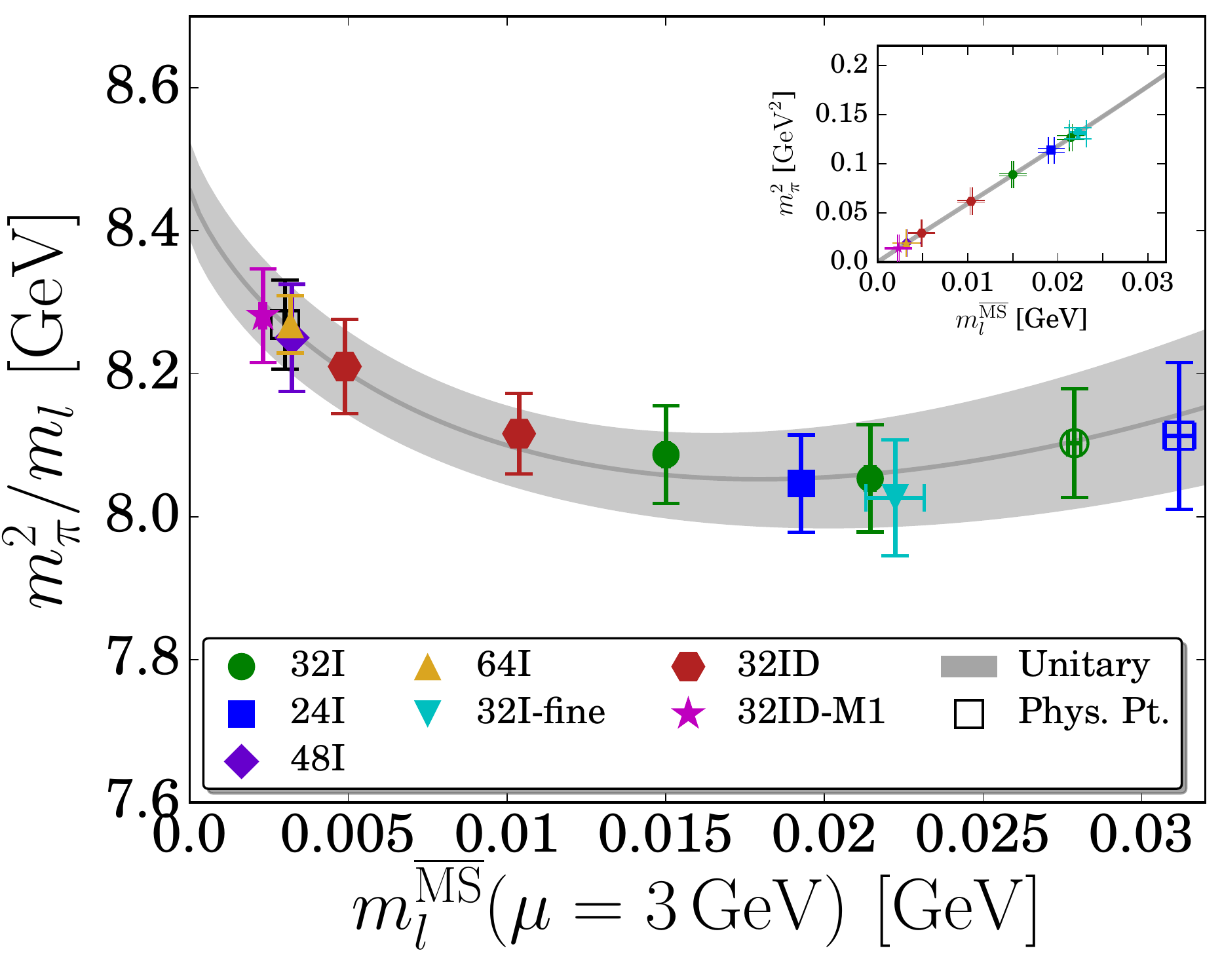}}
\subfloat[NNLO, 450 MeV cut]{\includegraphics[width=0.32\textwidth]{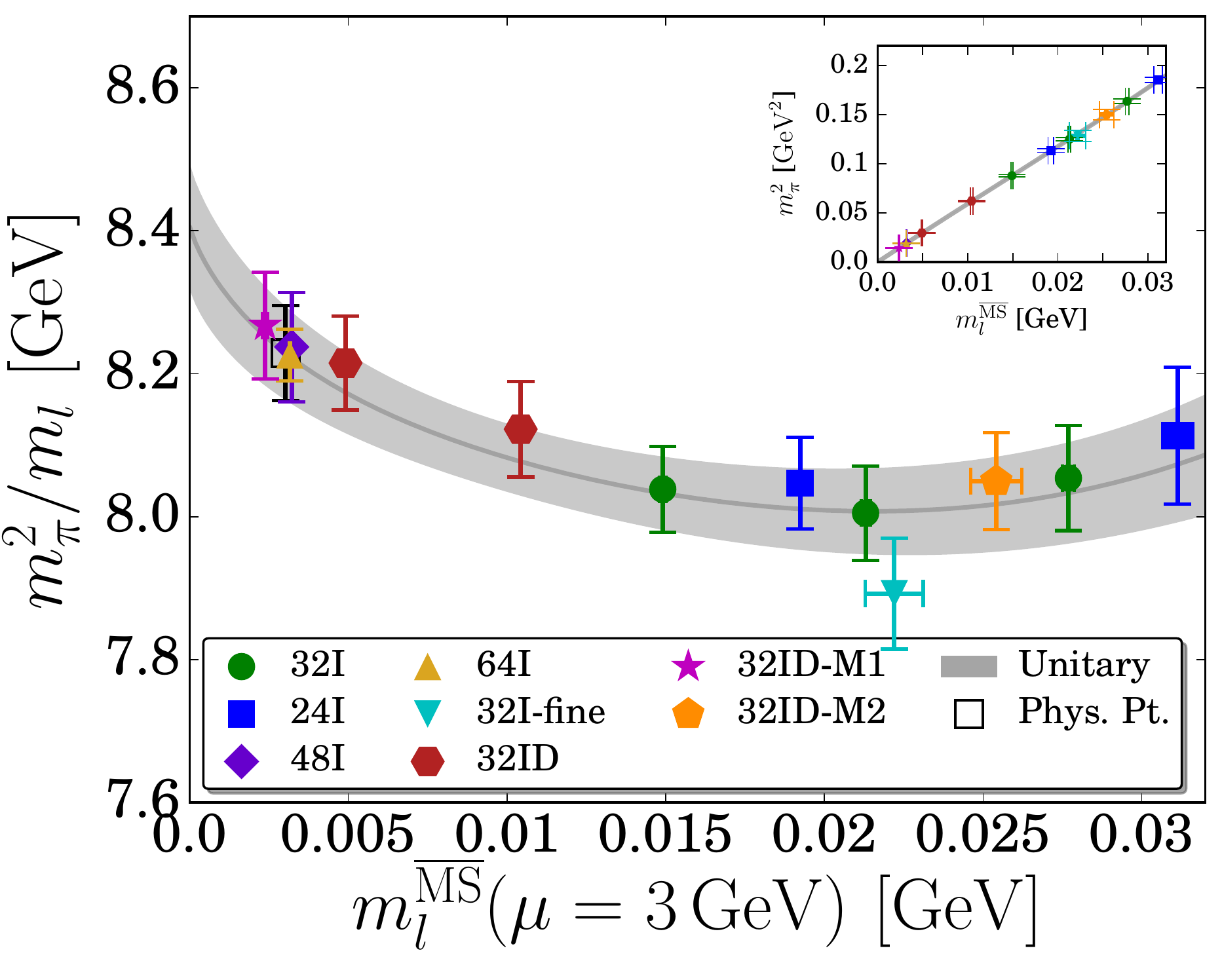}} 
\subfloat[NNLO, 450 MeV cut]{\includegraphics[width=0.32\textwidth]{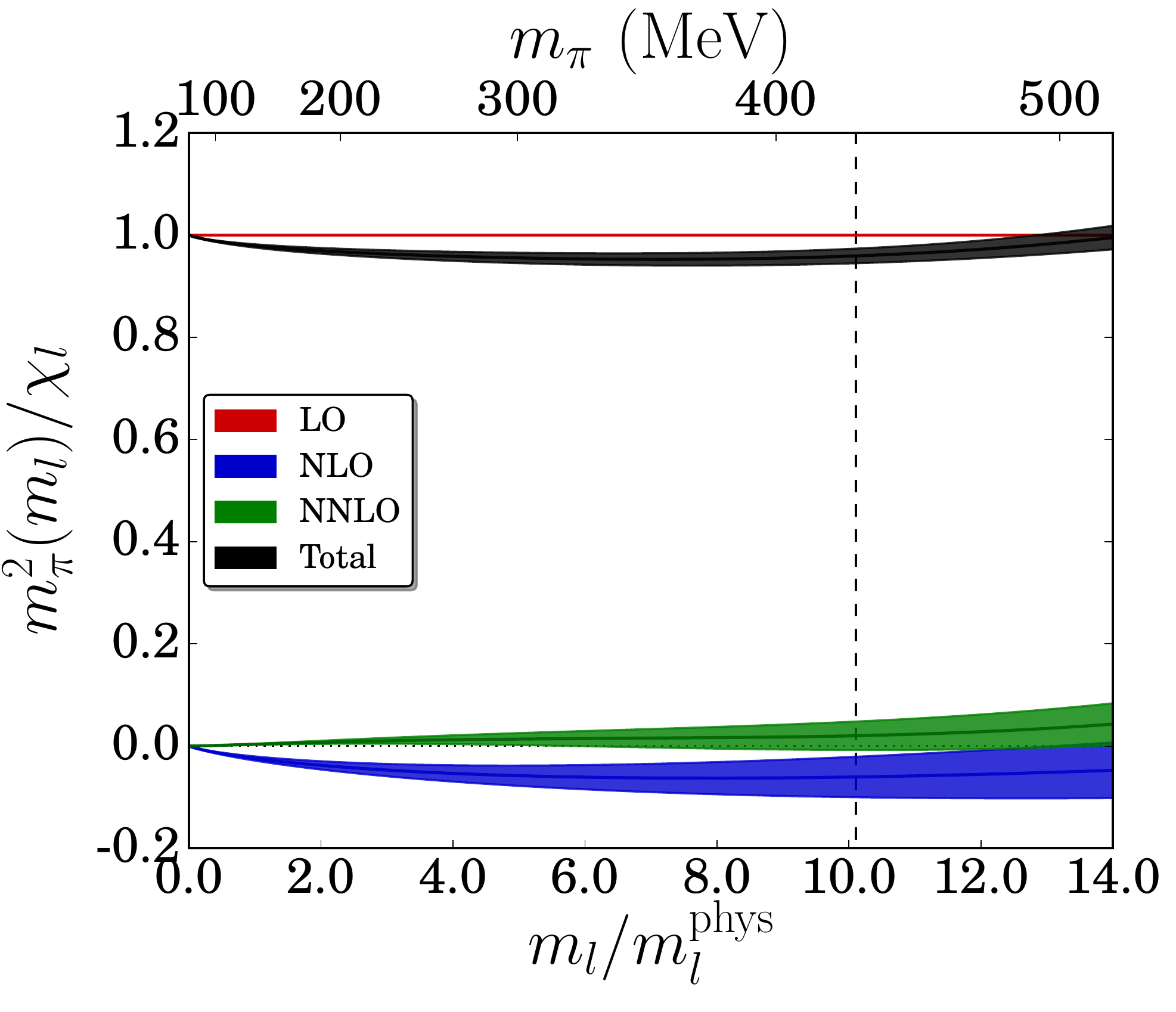}} \\
\subfloat[NLO, 370 MeV cut]{\includegraphics[width=0.32\textwidth]{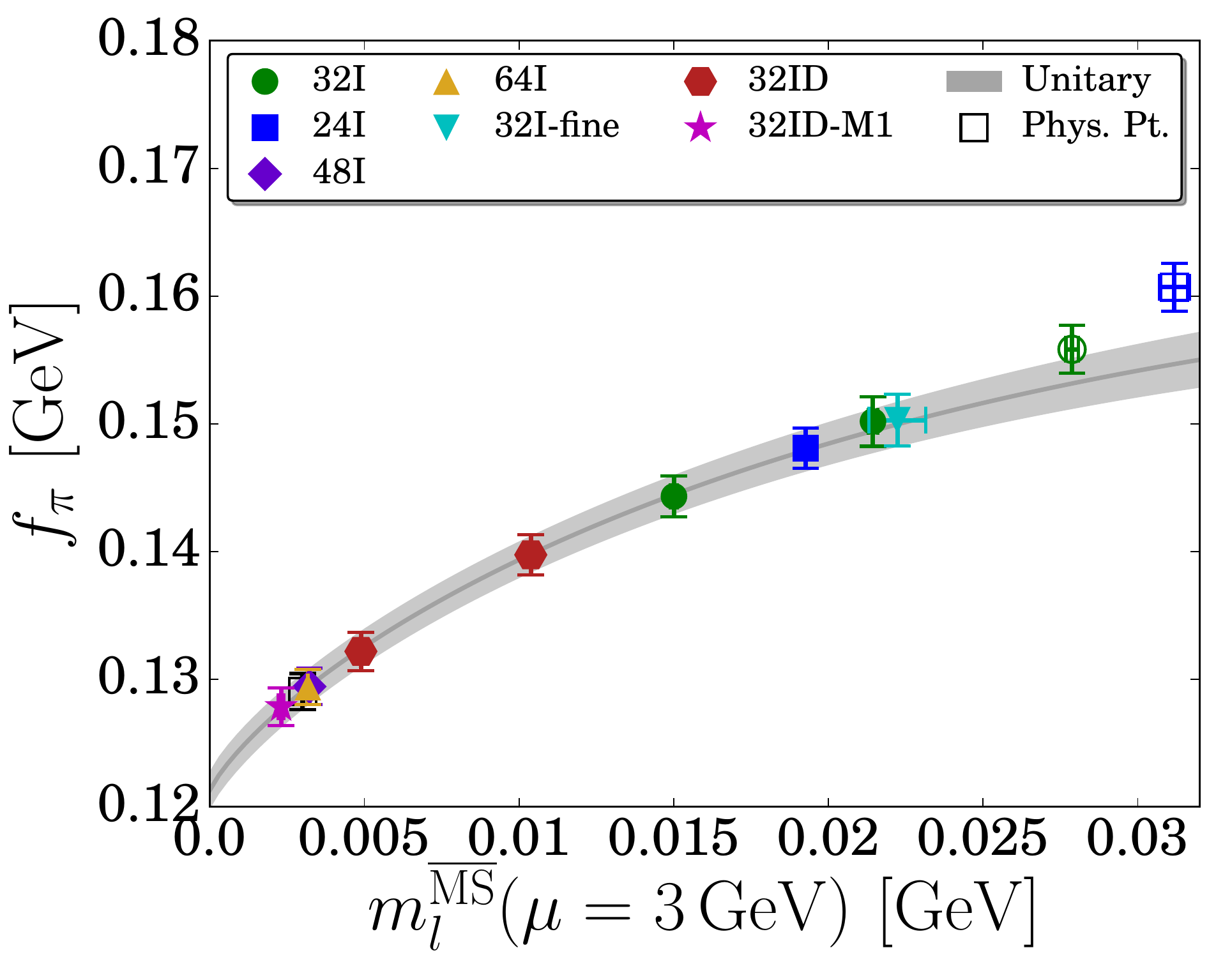}}
\subfloat[NNLO, 450 MeV cut]{\includegraphics[width=0.32\textwidth]{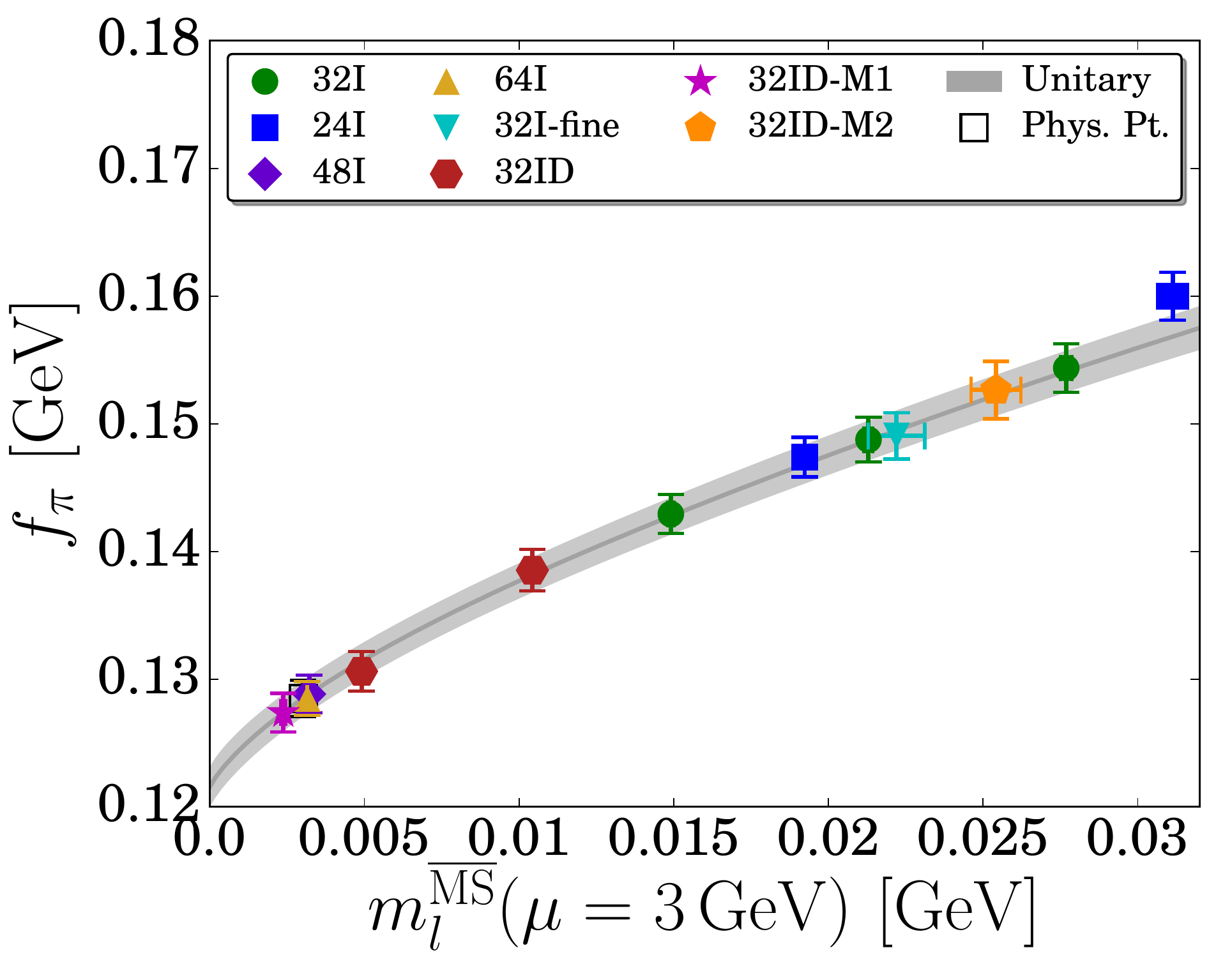}}
\subfloat[NNLO, 450 MeV cut]{\includegraphics[width=0.32\textwidth]{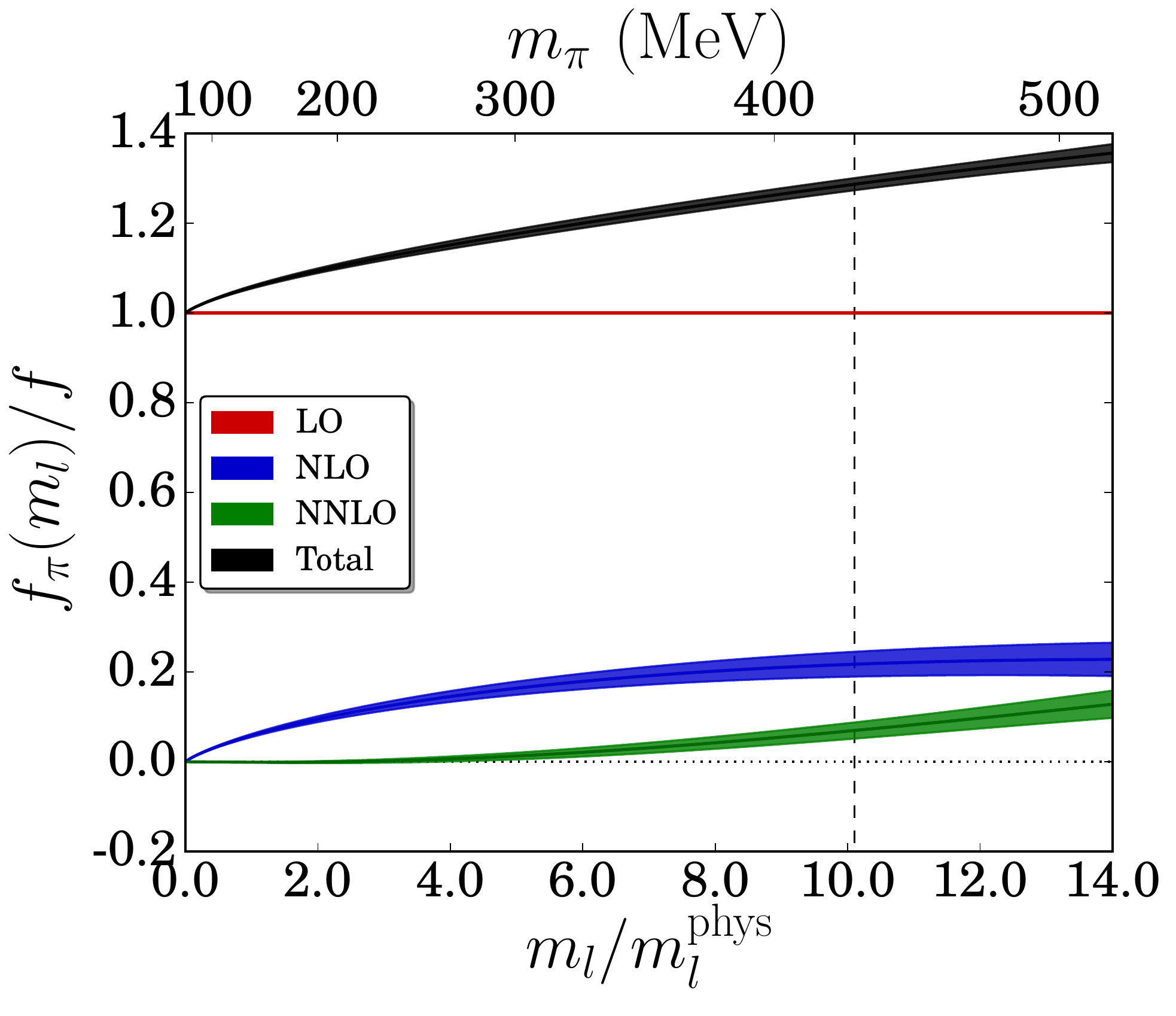}}
\caption{Left and middle columns: Chiral extrapolation of unitary measurements. The fit has been used to correct each data point to the physical strange quark mass and to take the infinite volume and continuum limits. Open symbols are data which is excluded from the fit. Right column: Decomposition of the $SU(2)$ expansion normalized by LO, where the vertical line marks the fit cut.}
\label{fig:unitary_expansion}
\end{figure}

In Figure \ref{fig:unitary_expansion} we overlay the unitary measurements of $m_{\pi}^{2}$ and $f_{\pi}$ on each ensemble with the $\chi \mathrm{PT}$ prediction obtained using the LECs from each fit. We observe that the NLO and NNLO fits to $m_{\pi}^{2}$ are completely consistent within statistics, however, we attribute this to the linearity of $m_{\pi}^{2}$ in the light quark mass: $\chi \mathrm{PT}$ predicts this linearity at tree-level, and thus the fits can easily match the full range of data within present statistical errors simply by keeping the corrections to leading order small. The fits to $f_{\pi}$ provide a more stringent constraint on the loop corrections, and indeed one can see a clear tension between the NLO fit and the heaviest 24I and 32I data, which is mitigated when the NNLO corrections are included. We interpret this as an indication that NLO $\chi \mathrm{PT}$ and our lattice data begin to systematically disagree at a scale of roughly $m_{\pi} \sim 350~\mathrm{MeV}$.

Since the $\chi \mathrm{PT}$ expansion is, in general, an asymptotic rather than convergent series, it is not obvious \textit{a priori} at what range of quark masses a given truncation of the $\chi \mathrm{PT}$ expansion should agree with QCD to a given precision. We explore this issue for the NNLO expansion in the rightmost column of Figure \ref{fig:unitary_expansion} by plotting the decomposition into LO, NLO, and NNLO terms, normalized by LO, using the LECs from the NNLO fit with a 450 MeV cut. At the physical light quark mass we observe a nicely ordered series with
\begin{equation}
\begin{split}
m_{\pi}^{2}/m_{l} &= 1.0000 - 0.0245(41) + 0.0034(10) \\
f_{\pi}/f &= 1.0000 + 0.0586(35) - 0.0011(7)
\end{split},
\end{equation}
suggesting that NLO corrections are $\mathcal{O}(2-5 \%)$ and NNLO corrections are $\mathcal{O}(0.1-0.3 \%)$ relative to LO. For heavier quark masses, the most obvious sign of distress in the expansion observed in Figure \ref{fig:unitary_expansion} is associated with $f_{\pi}$: at the physical light quark mass one has $\rm |NNLO| << |NLO|$, but the NNLO corrections grow relative to NLO as $m_{l}$ is increased. We find, for example, that $\rm |NNLO| \gtrsim 0.5 |NLO|$ within statistical error when $m_{\pi} \gtrsim 450~\mathrm{MeV}$, suggesting that the NNLO $SU(2)$ expansion starts to become unreliable at a scale $m_{\pi} \sim 450~\mathrm{MeV}$.

\begin{figure}[!ht]
\vspace{-0.15cm}
\centering
\subfloat{\includegraphics[width=0.9\textwidth]{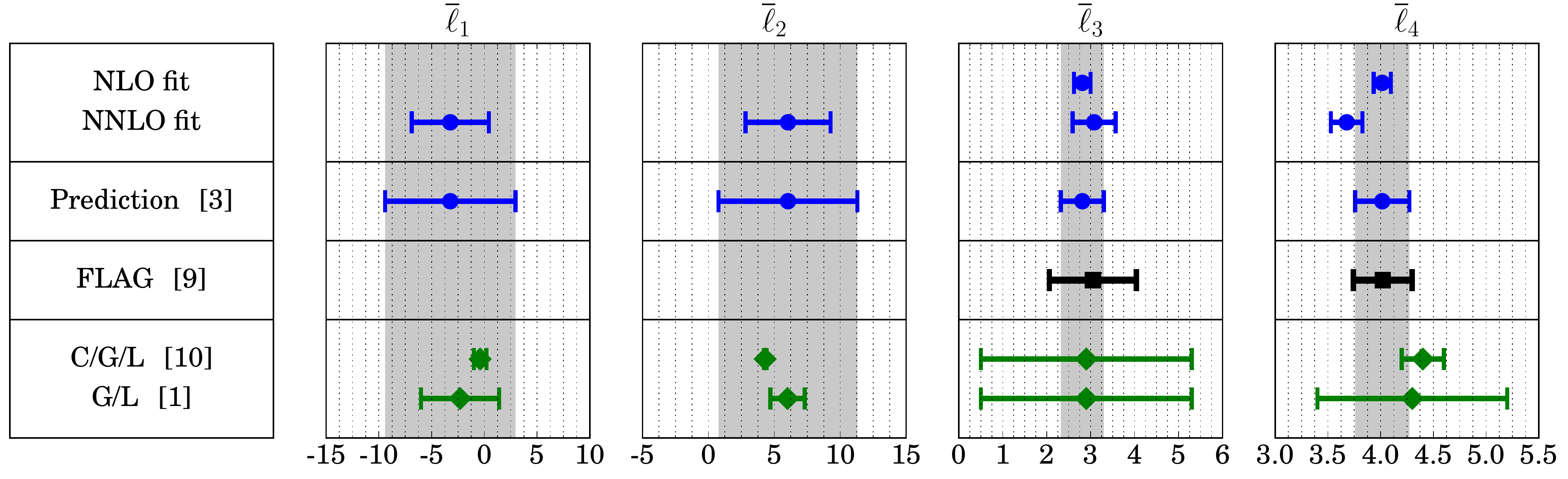}}\\
\subfloat{
	\resizebox{0.49\columnwidth}{!}{
		\begin{tabular}{c|c|c}
			\hline
			\hline
			\rule{0cm}{0.4cm}LEC & NLO fit ($370 \, \mathrm{MeV}$ cut) & NNLO fit ($450 \, \mathrm{MeV}$ cut) \\
			\hline
			\rule{0cm}{0.4cm}$\overline{\ell}_{1}$ & --- & -3.2(3.7) \\
			$\overline{\ell}_{2}$ & --- & 6.0(3.2) \\
			$\overline{\ell}_{3}$ & 2.81(19) & 3.08(49) \\
			$\overline{\ell}_{4}$ & 4.015(81) & 3.68(15) \\
			$10^{3} l_{7}$ & --- & 6.5(3.8) \\
			\hline
			\hline
		\end{tabular}
	}
}
\subfloat{
	\resizebox{0.49\columnwidth}{!}{
		\begin{tabular}{c|c|c}
			\hline
			\hline
			& Prediction & Expt. \cite{BlochDevaux:2009zzb} \\
			\hline
			\rule{0cm}{0.4cm}$m_{\pi} a_{0}^{0}$ & 0.199(9) & 0.221(5) \\
			$m_{\pi} a_{0}^{2}$ & -0.040(3) & -0.043(5) \\
			$\displaystyle \left[ \frac{\left( m^{2}_{\pi^{\pm}} - m^{2}_{\pi^{0}} \right)}{\left( m_{d} - m_{u} \right)^{2}} \right]_{\rm QCD}$ & 31(18) & --- \\
			\hline
			\hline
		\end{tabular}
	}
}
\caption{Top row: NLO $SU(2)$ $\chi \mathrm{PT}$ LECs compared to the 2013 FLAG lattice averages and two phenomenological determinations. Bottom row: values for the NLO LECs and some additional one-loop $SU(2)$ predictions from the NNLO fit (right). The reported errors are statistical only.}
\label{fig:unquenched_lecs}
\end{figure}

In Figure~\ref{fig:unquenched_lecs} we compute the unquenched NLO $SU(2)$ LECs $\{ \overline{\ell}_{i} \}$ from the values of the partially quenched NLO $SU(2)$ LECs $\{ \hat{L}_{i}^{(2)} \}$ determined by our fits, and compare our results (blue circles) to the 2013 $N_{f} = 2 + 1$ FLAG lattice averages (black squares) 
\cite{FLAG2013} 
and two phenomenological fits (green diamonds) \cite{Gasser:1983yg, Colangelo:2001df}. We also plot our final predictions from Ref.~\cite{su2_chpt_paper}, which include a full systematic error budget summed in quadrature. We generally observe excellent consistency between our fits, and find that our results for $\overline{\ell}_{3}$ and $\overline{\ell}_{4}$ compare favorably with the FLAG averages and phenomenological fits. In our NNLO fits we are also able to constrain $\overline{\ell}_{1}$, $\overline{\ell}_{2}$, and the scale-independent NLO LEC $l_{7}$: while our results for $\overline{\ell}_{1}$ and $\overline{\ell}_{2}$ are consistent with the phenomenological
results, these LECs are determined more precisely by the $\pi \pi$ scattering-based phenomenological fits. As a final test of our results we make one-loop predictions for the $I=0$ and $I=2$ $\pi \pi$ scattering lengths ($a_{0}^{0},a_{0}^{2}$), and for the $\pi^{\pm}-\pi^{0}$ mass splitting due to the up-down mass difference. We choose to write the latter in the dimensionless form $(m_{\pi^{\pm}}^{2} - m_{\pi^{0}}^{2})/(m_{d}-m_{u})^{2}$.

\begin{figure}[!ht]
\vspace{-0.6cm}
\centering
\subfloat{\raisebox{1mm}{\includegraphics[width=0.5\textwidth]{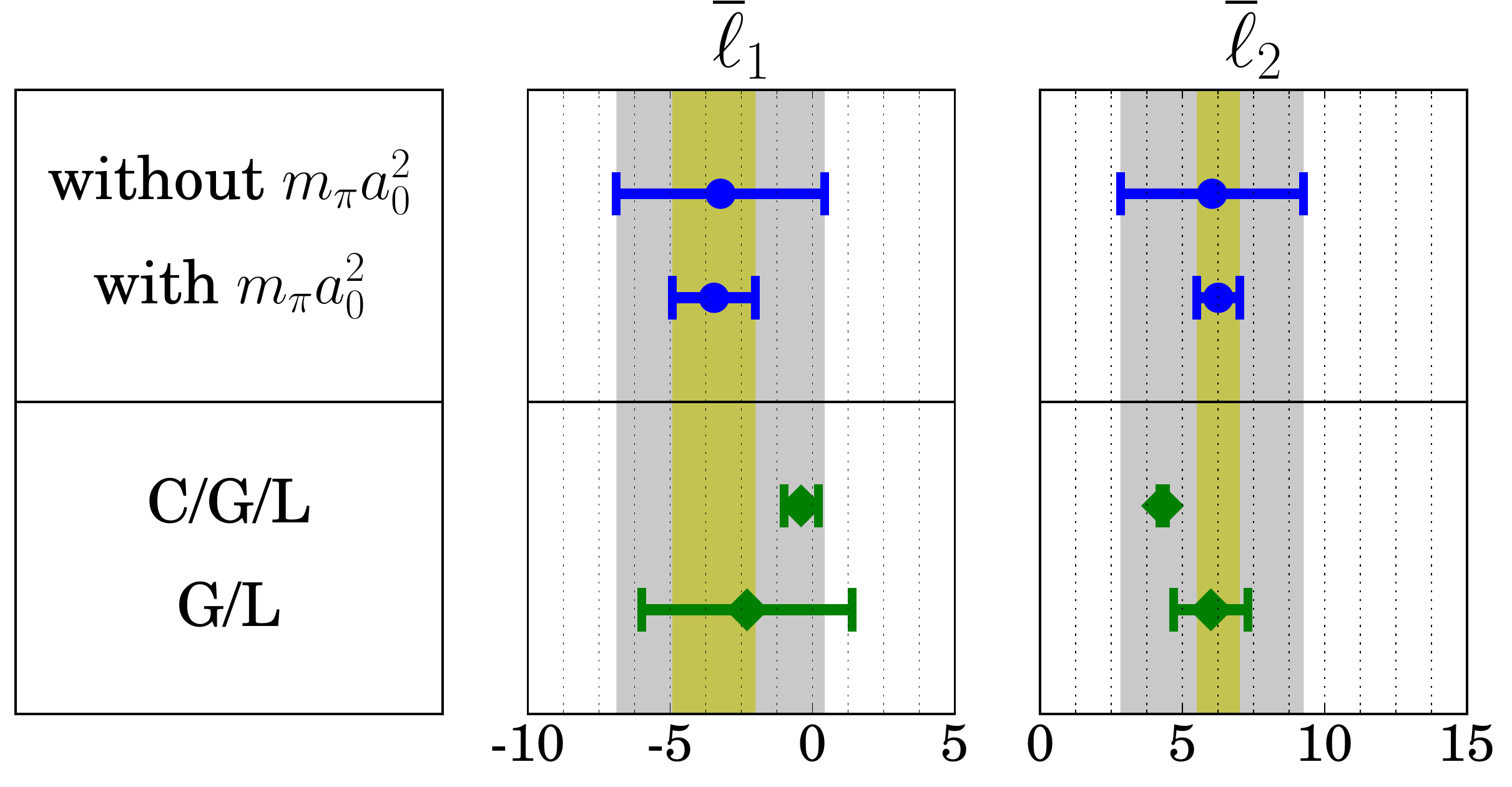}}}
\subfloat{\includegraphics[width=0.32\textwidth]{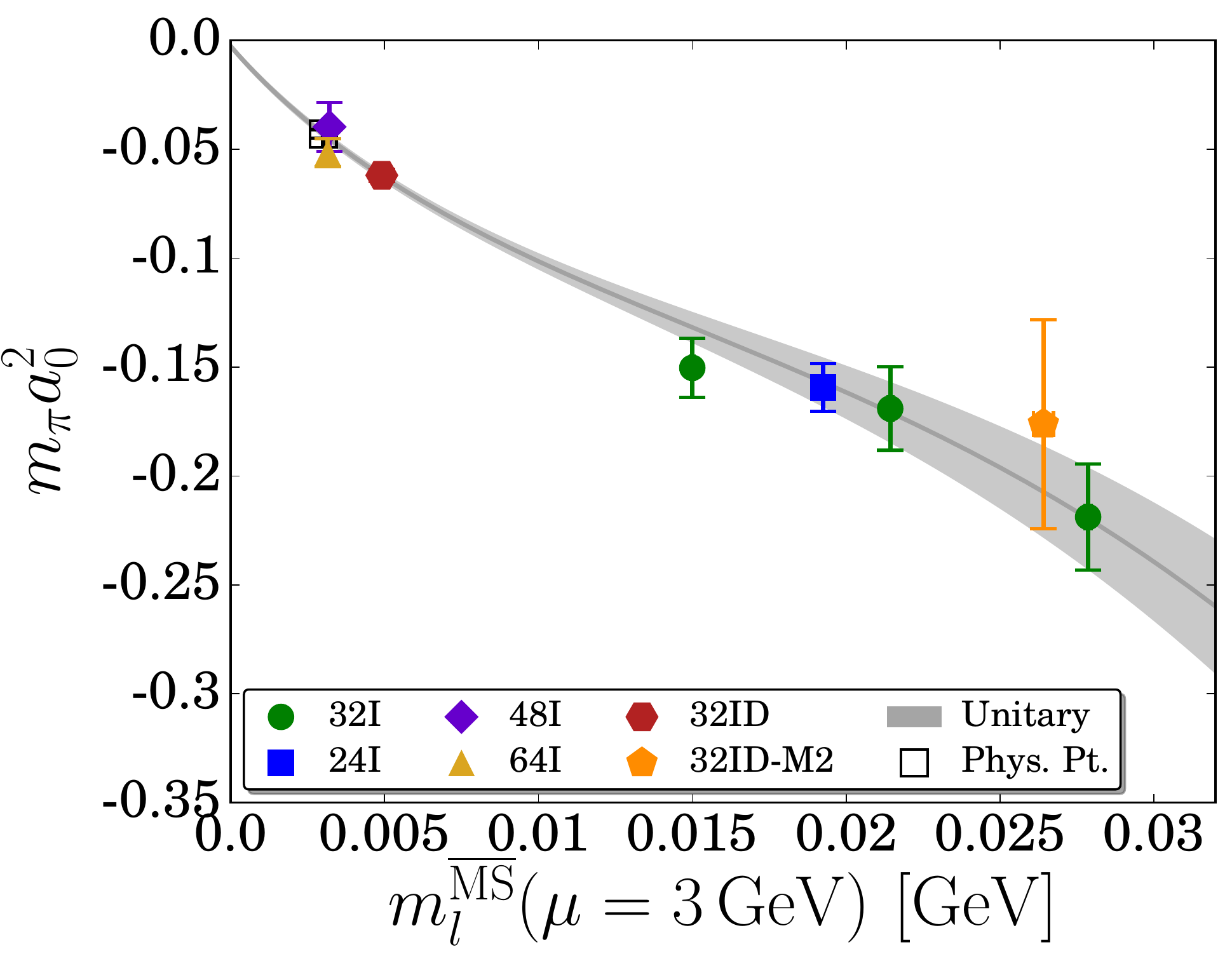}}
\caption{Preliminary results including the $I=2$ $\pi \pi$ scattering length ($a_{0}^{2}$). In the right panel the fit has been used to correct the data to $m_{s}^{\rm phys}$, and take the infinite volume and continuum limits.}
\label{fig:pipi_scatt}
\end{figure}

As an extension of this work, we have begun to explore fits which also include results for the $I=2$ $\pi \pi$ scattering length $a_{0}^{2}$, which directly constrains $\overline{\ell}_{1}$ and $\overline{\ell}_{2}$ at NLO in $\chi \mathrm{PT}$. In Figure \ref{fig:pipi_scatt} we repeat our NNLO fit with a 450 MeV pion mass cut, and include measurements of the $I=2$ $\pi \pi$ scattering length ($a_{0}^{2}$) on a subset of the ensembles. We use NLO $\chi \mathrm{PT}$ supplemented with terms linear in $m_{h}$ and $a^{2}$ as the chiral ansatz for $m_{\pi} a_{0}^{2}$. We find that the values of $\overline{\ell}_{1}$ and $\overline{\ell}_{2}$ we obtain are consistent with our earlier fit, but with substantially improved statistical resolution.

\section{Conclusions}

In this work we have performed fits of pseudoscalar masses and decay constants from a series of RBC-UKQCD domain wall fermion ensembles to the corresponding formulae in NNLO $SU(2)$ $\mathrm{PQ} \chi \mathrm{PT}$. We have reported values for a large set of partially quenched LECs, and used these values to compute NLO unquenched LECs which we compare to other results in the literature. We find that the chiral fits are generally of excellent quality and match our lattice data with percent-scale accuracy provided the mass cut is chosen appropriately.

Future work will incorporate calculations of the $I = 2$ $\pi \pi$ scattering length and the pion vector form factor. Including these results in our chiral fits will give first-principles determinations of the scattering length $a_{0}^{2}$, the pion charge radius $\langle r^{2} \rangle_{V}^{\pi}$, and the $SU(2)$ LEC $\overline{\ell}_{6}$, as well as sharpen the predictions for $\overline{\ell}_{1}$ and $\overline{\ell}_{2}$, which are currently determined most precisely by phenomenological fits to experimental data.

\end{document}